\documentstyle[aps,12pt]{revtex}

\begin{document}
\draft
\author{V. V. ULyanov and O. B. Zaslavskii$^{1}$}
\address{Department of Physics, Kharkov State University, Svobody Sq.4, Kharkov\\
310077, Ukraine\\
$^{1}$E-mail: aptm@kharkov.ua}
\title{Tunnelling series in terms of perturbation theory for quantum spin systems}
\maketitle

\begin{abstract}
Considered is quantum tunnelling in anisotropic spin systems in a magnetic
field perpendicular to the anisotropy axis. In the domain of small field the
problem of calculating tunnelling splitting of energy levels is reduced to
constructing the perturbation series with degeneracy, the order of
degeneracy being proportional to a spin value. Partial summation of this
series taking into account ''dangereous terms'' with small denominators is
performed and the value of tunnelling splitting is calcaulated with
allowance for the first correction with respect to a magnetic field.
\end{abstract}


Tunnelling splitting of energy levels and the appearance of gaps in the
energy spectrum of a quantum system is connected with breaking the symmetry
of an unperturbed system by perturbation removing degeneracy. Usually, this
effect is realized in quantum mechanics in two typical situations. The first
one consists in that the Hamiltonian can be represented in the form $%
H=H_{0}+V$ where $H_{0}$ is the unperturbed Hamiltonian and $V$ is
perturbation, so correction to energy levels can be calculated with the help
of some version of the perturbation theory. Another situation is tunnelling
splitting of energy levels for a particle moving in a symmetric double well
due to tunnelling penetration into a classically forbidden region. In that
case the tunnelling splitting can be found by means of one of quasiclassical
methods (WKB, instanton calculus, etc.). Although the physical reason -
removing degeneracy - is the same in both cases, it is described in terms of
essentially different languages regions of applicability of which, generally
speaking, do not match.

It turns out, however, that there exists a whole class of physically
interesting systems for which both approaches can be combined - anisotropic
spin systems whose Hamiltonian is a quadratic-linear combination of spin
operators. Their energy spectrum admits rigorous description in terms of an
effective potential \cite{rep}: energy levels of a spin system coincide with 
$2S+1$ low-lying levels for a particle moving in a potential field of a
certain form. If such a spin system is of the easy-type, for a sufficiently
small field the corresponding potential has the form of a double well
typical of tunnelling effects, so the energy splitting can be obtained, for
example, with the help of instanton formulas \cite{rep}, \cite{enz} accuracy
of which grows as a magnetic field decreases and the spin value increases.
On the other hand, in the domain of small fields the perturbation theory can
be applied. Thus, for sufficiently small magnetic fields and $S\gg 1$ both
approaches match.

The nontrivial peculiarly of the perturbation theory in such a situation
consists in that the order of degeneracy is proportional to the spin value 
\cite{zh}. As the quantity $S$ plays the role of the inverse Planck constant 
$\hbar $, it turns out that the degree of degeneracy tends to infinity in
the quasiclassical limit. In so doing, the terms like $B^{2S}$ appear in the
perturbation series which are nonanalytic with respect to $S^{-1}$in the
limit $S\rightarrow \infty $ that in itself testifies that, although
calculations are carried out on the basis of perturbation theory, the effect
in fact admits the tunnel interpretation.

In the leading order of perturbation theory the result of the calculation of
tunnelling splitting has been reported in \cite{gar} (without detailed
derivation) and, independently, in \cite{rep} where a consistent derivation
of the formulas for energy splitting was suggested. This derivation was not
direct application of standard formulas of the perturbation theory form
textbooks and required careful singling out and summing all ''dangerous''
terms. In the present paper we perform the next step and calculate the first
non-vanishing correction to the tunnelling splitting for an arbitrary value
of a spin. The approach developed in \cite{rep} and the present paper is of
systematic character that enables one to construct the series with respect
to a magnetic field value including high-order corrections. In this sense
the calculation of tunnelling rate is reduced to summing up the perturbation
theory series with a given accuracy.

In recent years the interest to the nature of quantum tunnelling in spin
systems of different nature sharply increased. It concerns the crossover
between quantum and thermal domains \cite{cros}, tunnelling in small
ferromagnet particles \cite{part}, many-spin molecules \cite{mol}, etc.
Therefore, formulae for the tunnelling splitting obtained below can be of
interest not only from a viewpoint of general methods of quantum theory but
for description of concrete physical systems observable in experiments.

Let us consider the system described by the Hamiltonian 
\begin{equation}
H=-S_{z}^{2}-BS_{x}  \label{ham}
\end{equation}
Here $S_{i}$ are the operator of a corresponding spin component, $B$ is a
magnetic field. Since the first terms enters (\ref{ham}) with a negative
sign, the system has an easy axis $z$. In the absence of a magnetic field
all energy levels, except the one with $S_{z}=0$ in the case of an integer $%
S $, are two-fold degenerate. A magnetic fields removes this degeneracy..
The problem is to calculate the energy splitting due to quantum tunnelling
with the first correction in $B$ taken into account when $B\rightarrow 0$.

Consider the Brillouin-Wigner perturbation series 
\begin{equation}
E=\varepsilon _{\sigma }+\frac{V_{\sigma ,\sigma -1}^{2}}{\alpha _{\sigma -1}%
}+\frac{V_{\sigma ,\sigma +1}^{2}}{\alpha _{\sigma +1}}...  \label{bril}
\end{equation}
where $\varepsilon _{\sigma }=-\sigma ^{2}$ corresponds to the unperturbed
level with a $z$-projection of spin equal to $\sigma $, $\alpha _{\sigma
}=E-\varepsilon _{\sigma }$, the perturbation $V=-BS_{x}$ and we took into
account that diagonal matrix elements of perturbation are equal to zero. In
order to determine $\Delta E_{\sigma }$ correctly, one needs to take into
account explicitly the role of small denominators in ''dangerous'' terms of
the perturbation series by performing a partial summation of (\ref{bril}).
Following \cite{rep}, we sketch this procedure below. Consider, for
simplicity, the splitting for the ground energy level $\Delta E_{0}$. The
leading ''dangerous'' terms $f$ reads 
\begin{equation}
f=\frac{(V_{S,S-1}V_{S-1,S-2}...V_{-S+1,-S})^{2}}{[(E-\varepsilon
_{S-1})(E-\varepsilon _{S-2})...(E-\varepsilon _{-S+1})]^{2}}(E-\varepsilon
_{-S})^{-1}
\end{equation}
If we introduce a sequence of points on a number axis from $S$ to $-S$
corresponding to different values of $\sigma $, then this term corresponds
to a single sequential transition from $S$ to $-S$ and back. Now take into
account that some elements of the number axis cab ne passed back and forward
a few times (after a transition from $S$ to $-S$ the system returns to the
intermediate point $\sigma $, then to $-S$ again and so on, passing the
intermediate values of $\sigma $). This suggests that additional factors
appear in the perturbation series. Thus in the sum over terms which contain $%
f$ as a common factor one should include higher orders of the perturbation.
It is easy to guess that this leads to the following form of the last
retained term in the perturbation series: 
\begin{equation}
f\rightarrow \tilde{f}=f(1+r+r^{2}+...)=f(1-r)^{-1}\text{, }r=(E-\varepsilon
_{S})^{-1}\chi \text{, }\chi =\frac{V_{-S,-S+1}V_{-S+1,-S}}{E-\varepsilon
_{-S+1}}+...  \label{2f}
\end{equation}
The factor $\chi $ has the same form as the initial perturbation series (\ref
{bril}) that allows one to construct a simple equation with respect to $E$: 
\begin{equation}
(E-\varepsilon _{\sigma }-\frac{V_{S,S-1}^{2}}{E-\varepsilon _{S-1}}-\frac{%
V_{S,S+1}^{2}}{E-\varepsilon _{S+1}}+...)^{2}=g_{S}^{2}\text{, }g_{S}=\frac{%
V_{S,S-1}V_{S-1,S-2}...V_{-S+1,-S}}{(E-\varepsilon _{S-1})(E-\varepsilon
_{S-2})...(E-\varepsilon _{-S+1})}  \label{e0}
\end{equation}

Generalization to other levels is straightforward and the result reads 
\begin{equation}
(E-\varepsilon _{\sigma }-\frac{V_{\sigma ,\sigma -1}^{2}}{E-\varepsilon
_{\sigma -1}}-\frac{V_{\sigma ,\sigma +1}^{2}}{E-\varepsilon _{\sigma +1}}%
+...)^{2}=g_{\sigma }^{2}\text{, }g_{\sigma }=\frac{\Pi _{k=0}^{2\sigma
-1}V_{\sigma -k,\sigma -k-1}}{\Pi _{k=1}^{2\sigma -1}\alpha _{\sigma -k}}
\label{eq}
\end{equation}
The differences of the signs corresponds to the level splitting (the
remaining terms coincide, making no contribution to the splitting). Taking
advantage of the explicit form of the matrix elements $V_{\sigma ,\sigma
^{\prime }}$ we get 
\begin{equation}
\Delta ^{(0)}E_{\sigma }=2g_{\sigma }=2(\frac{B}{2})^{2\sigma }\frac{(\sigma
+S)!}{(S-\sigma )!(2\sigma -1)!}  \label{de}
\end{equation}
in accordance with \cite{gar}, \cite{rep} where $\sigma =S-n$, $n$ is the
number of the unperturbed level.

The first correction to (\ref{de}) stems from three origins. First, the
value of $g_{\sigma }$ in (\ref{eq}) acquires the correction due to field
dependence of levels $E_{\sigma }$. Second, the correction appears due to $%
E_{\sigma }$ which enter denominators in terms containing $V^{2}$ in the
left hand side of (\ref{eq}). Third, the correction arises because of the
next order terms in the perturbation series itself which change the
structure of $g_{\sigma }$. Let us write down the total fractional
correction as $\xi \equiv $ $\frac{\Delta E_{\sigma }}{\Delta E_{\sigma
}^{(0)}}\equiv \xi _{1}+\xi _{2}+\xi _{3}$ and calculate each term
separately.

The first term $\xi _{1}=\Pi _{k=1}^{2\sigma -1}\frac{(E^{(0)}-\varepsilon
_{\sigma -k})}{(E-\varepsilon _{\sigma -k})}$. Substituting into this
formula the explicit expression for unperturbed levels $E_{\sigma
}^{(0)}=-\sigma ^{2}$ and take into account the second order correction to $%
E $ which can be easily found from the standard perturbation theory we
obtain ($n=S-\sigma $) 
\begin{equation}
\xi _{1}=-\frac{B^{2}[n^{2}-2Sn+S(2S+1)]}{(2S-2n)[(2S-2n)^{2}-1]}\Sigma
_{k=1}^{2S-2n-1}k^{-1}  \label{f}
\end{equation}

The second type contribution to the correction can be found directly from
the Brillouin-Wigner perturbation series by elementary methods: 
\begin{equation}
\xi _{2}=-\frac{B^{2}}{4}[\frac{(n+1)(2S-n)}{(2S-2n-1)^{2}}+\frac{n(2S-n+1)}{%
(2S-2n+1)^{2}}]  \label{s}
\end{equation}

The calculation of $\xi _{3}$ is the nontrivial and rather cumbersome part
of computations. This quantity is connected with the correction to the
perturbation theory itself which is to be now cut off at terms of the order $%
B^{2S+2}$. According to \cite{rep}, a typical term of the perturbation
theory series arises as follows. It is necessary to take the segment $%
[-\sigma ,\sigma ]$ and allow for all possible paths of the transition
between these two points provided each an intermediate point on each path is
being passed two times (to and fro). The main corrections to these terms
arise if the transition from a given point to a next one contains one extra
jump to and one - fro (delay and ''marking time''). Summing up over all
possible points - origins of extra jumps to and fro - we obtain 
\begin{equation}
\xi _{3}=\Sigma _{k=2}^{2S-1}\frac{V_{\sigma -k,\sigma -k+1}^{2}}{\alpha
_{\sigma -k}\alpha _{\sigma -k+1}}  \label{t}
\end{equation}
Summing up all three contributions we find, after direct but somewhat
lengthy calculations, the final expression which proves to be surprisingly
simple: 
\begin{eqnarray}
\xi  &=&1-B^{2}\gamma \text{, }  \label{tot} \\
\gamma  &=&\frac{(2S+1)^{2}(\sigma +1)}{2(2\sigma -1)^{2}(2\sigma +1)^{2}} 
\nonumber
\end{eqnarray}

Thus, for an arbitrary energy gap (except the highest one in the case of
semi-integer $S$) we get 
\begin{equation}
\Delta E_{n}=\frac{(2S-n)!}{2^{2S-2n-1}n!(2S-2n-1)!^{2}}B^{2S-2n}[1-\frac{%
(2S+1)^{2}(S-n+1)}{2(2S-2n-1)^{2}(2S-2n+1)^{2}}B^{2}]  \label{result}
\end{equation}
This formula is supplemented for semi-integer $S$ by the expression for the
highest gap: 
\begin{equation}
\Delta E_{S-1/2}=(S+\frac{1}{2})B[1-\frac{1}{16}(S+\frac{3}{2})(S-\frac{1}{2}%
)B^{2}]  \label{sup}
\end{equation}

The obtained formulae, as was mentioned above, are valid when $B\rightarrow
0 $. However, the presence of the correction term in (\ref{result}) and (\ref
{sup}) aids to make the area of applicability of these formulas more
precise. First of all, the correction gives a qualitative notion about the
approximation: it is seen that without account for such a correction the
result is overstated. Further, the quantitative role of the correction turns
out two-fold.

First, it enables us to estimate the accuracy of the first approximation.
Let us restrict ourselves by remarks concerning the ground gap ($S>1/2$) 
\begin{equation}
\Delta E_{0}=\frac{S^{2}}{2^{2S-3}(2S)!}B^{2S}[1-\frac{(S+1)}{2(2S-1)^{2}}%
B^{2}]  \label{ground}
\end{equation}
It is seen from this formula that in the case $S\gg 1$ ,with the correction
neglected, the fractional error is $B^{2}/8S$. That means that even for $B=%
\sqrt{S}$ the computation of the ground gap gives the error approximately $%
10\%$ while in the case $B=1$ it proves to be very small, being of the order 
$0.1/S$.

Second, the allowance for the correction essential increases the accuracy of
computations. Thus, even for so large values of a magnetic field as $\sqrt{S}
$ can be thought of to be, taking into account the correction enables one to
make the fractional error one order smaller.

It is worth noting that while computing energy gaps numerically even for
moderately large values of $S$ one is led to take into account very big
number of digits that can be achieved only by modern computer computation
systems. For example, in the case $B=1$, $S=10$ the gap has the order $%
10^{-20}$ while for $S=50$ it is of the order $10^{-200}$.

Either the main term or the correction in (\ref{ground}) are in agreement
with the instanton approach \cite{rep} but for $S\gg 1$. Meanwhile, it is
worth stressing that formulae (\ref{result})-(\ref{sup}) are applicable for
all values of $S\,$and embrace all the gaps.

In the case of an easy-plane spin system, i.e. a system with a Hamiltonian $%
H=+S_{z}^{2}-BS_{x}$, the energy spectrum of which differs from that of the
easy-axis type by the interchange of the sign of energy. Therefore, formulae
for gaps remains valid provided they are renumbered, so that the lowest gap
becomes the highest and vice versa. As far as the effective potential is
concerned, in this case we have the model of a periodic potential energy in
which the energy levels of a spin system coincide with the corresponding
edges of energy bands (\cite{rep}).

The elaborated approach describes quantum tunnelling in terms of the
perturbation theory and enables one, in principle, to construct the energy
level splitting as a series with respect to $B$ with any given accuracy.

\section{acknowledgment}

\begin{description}
\item  The work of O. Z. was supported by International Science Education
Program (ISEP), grant No. QSU082068.
\end{description}





%
%

%
%

\end{document}